\documentclass[twocolumn, pra, superscriptaddress]{revtex4-1}
\usepackage{amssymb}
\usepackage{amsmath}
\usepackage{epsfig}
\usepackage{color}
\usepackage{graphics, graphicx}
\usepackage{bbold}
\usepackage{psfrag}
\usepackage{mathcomp}
\usepackage{subfigure}
\usepackage{verbatim}
\usepackage{color}
\usepackage[colorlinks,citecolor=blue]{hyperref}

\begin{document}
\title{Interaction-induced Bloch Oscillation in a Harmonically Trapped and Fermionized Quantum Gas in One Dimension}
\author{Lijun Yang}
\affiliation{Key Laboratory of Quantum Information, University of Science and Technology of China,
CAS, Hefei, Anhui, 230026, People's Republic of China}
\affiliation{Synergetic Innovation Center of Quantum Information and Quantum Physics, University of Science and Technology of China, Hefei, Anhui 230026, China}
\author{Lihong Zhou}
\affiliation{Beijing National Laboratory for Condensed Matter Physics, Institute of Physics, Chinese Academy of Sciences, Beijing 100190, China}
\author{Wei Yi}
\affiliation{Key Laboratory of Quantum Information, University of Science and Technology of China,
CAS, Hefei, Anhui, 230026, People's Republic of China}
\affiliation{Synergetic Innovation Center of Quantum Information and Quantum Physics, University of Science and Technology of China, Hefei, Anhui 230026, China}
\author{Xiaoling Cui}
\email{xlcui@iphy.ac.cn}
\affiliation{Beijing National Laboratory for Condensed Matter Physics, Institute of Physics, Chinese Academy of Sciences, Beijing 100190, China}
\date{\today}

\begin{abstract}
Motivated by a recent experiment by F. Meinert {\it et al},
arxiv:1608.08200, we study the dynamics of an impurity moving in the
background of a harmonically trapped one-dimensional Bose gas in the
hard-core limit. We show that due to the hidden ``lattice" structure
of background bosons, the impurity effectively feels a
quasi-periodic potential via impurity-boson interactions that can
drive the Bloch oscillation under an external force, even in the absence of real lattice potentials.
Meanwhile, the inhomogeneous density of trapped bosons imposes an
additional harmonic potential to the impurity, resulting in a
similar oscillation dynamics but with a different period and
amplitude. We show that the sign and strength of the impurity-boson coupling
can significantly affect the above two potentials in determining the impurity dynamics.
\end{abstract}

\maketitle

\section{Introduction}

Bloch oscillation (BO) describes a striking quantum phenomenon, in which the motion of a particle under a periodic potential and an external force is oscillatory, rather than linear, as time evolves~\cite{Bloch, Zener}. It has been observed in semiconductor superlattices~\cite{BO_superlattice} and in cold atoms with optical lattices~\cite{BO_OL1,BO_OL2, BO_OL3, BO_OL4}. Physically, this phenomenon is due to the Bragg scattering of the particle at the edge of the Brillouin zone, which is supported by the translational invariance of lattice potentials. It is then interesting to ask the question: is translational invariance necessary for the occurrence of BO? A recent experiment at Innsbruck~\cite{expt} seems to suggest the answer {\it no}. In this experiment, oscillatory dynamics of an impurity interacting with hard-core bosons trapped in one dimension have been observed, in the absence of any periodic confinements. In Ref.~\cite{expt} and earlier theories~\cite{Gangardt1, Gangardt2, Gangardt3}, this phenomenon was attributed to the Bragg scattering with bosons at the edge of an emergent Brillouin zone, which causes the impurity momentum to change by twice the Fermi momentum of bosons with no energy cost.

\begin{figure}[t]
\includegraphics[width=8.5cm]{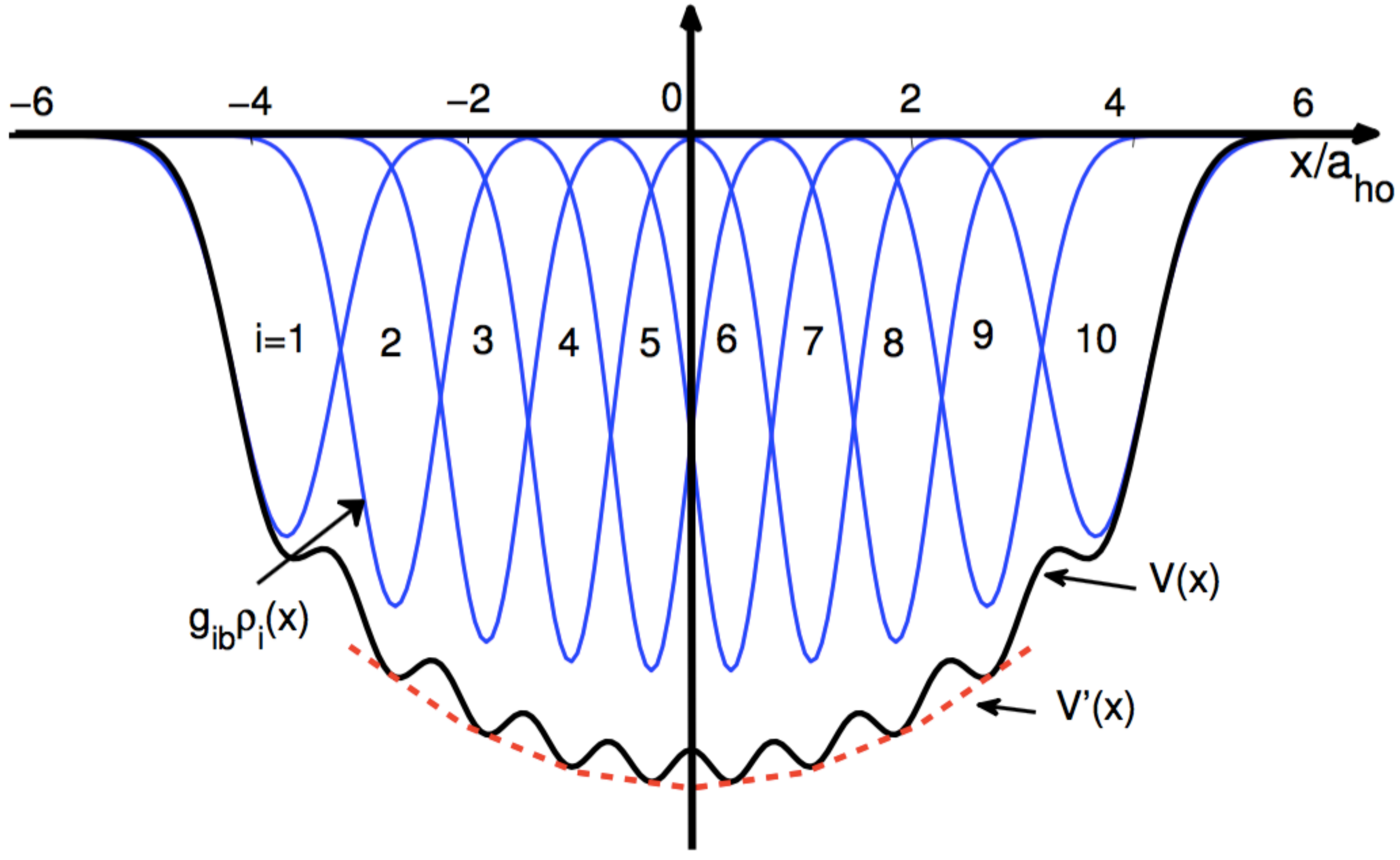}
\caption{(Color online). Schematic plot of the ``lattice" potential, $V(x)=g_{ib} \sum_i \rho_i(x)$, for an impurity moving in the background of N hard-core bosons. $\rho_i(x)$ ($i=1,2...N$) is the density distribution of the i-th ordered boson in the trap, and $g_{ib}$ is the impurity-boson coupling strength. In the plot we take $N=10$ and $g_{ib}<0$. $a_{ho}$ is the typical length of the harmonic confinement. The dashed line shows the deviation of $V(x)$ from ideal lattice potential due to the density inhomogeneity of bosons in a harmonic trap.} \label{fig1}
\end{figure}

In this work, we give an alternative interpretation of the oscillation dynamics observed in Ref.~\cite{expt}, by adopting the concept of effective spin chain in strongly interacting atomic gases in one dimensional (1D) traps~\cite{Santos, Zinner, Pu, Levinsen,Yang}. The idea of spin chain is based on the fact that for fermionized (or impenetrable) particles in one dimension, their spatial order is fixed, and the probability peak of finding the i-th ordered particle (see $\rho_i(x)$ in Fig.~\ref{fig1}) is well separated from those of the neighboring ordered particles. Therefore an underlying ``lattice" chain is automatically formed by mapping the order index to the corresponding site index~\cite{Yang}. 
In this way, various spin-chain Hamiltonians can be constructed in response to different external perturbations, including interactions, gauge fields and trapping potentials, which have been used to address and engineer novel spin spirals and magnetic orders in 1D systems~\cite{Santos, Zinner, Pu, Levinsen,Yang, Cui, Blume, Zinner2, Levinsen2, Cui2, Chen, Pu2, Santos2}.
Experimentally, an anti-ferromagnetic spin chain has been recently confirmed in a small cluster of 1D trapped fermions~\cite{Jochim_expt}.

Now we apply the idea of spin chain to the Innsbruck experiment~\cite{expt}. Note that the ``spin chain" here refers to the ordered (hardcore) bosons in the coordinate space. As the impurity-boson interaction can be converted to the impurity interacting with all the ordered particles (see Fig.~\ref{fig1}), it then becomes clear that the impurity should effectively feel a ``lattice" potential originating from the ``lattice" structure of ordered bosons. Such a ``lattice" potential is quasi-periodic, with the lattice spacing approximately the inter-particle distance of bosons and the lattice depth proportional to the impurity-boson coupling strength. This lattice potential then gives rise to the BO of impurity under an external force. This picture, compared to that in Ref.~\cite{expt}, provides more information on the role of impurity-boson interaction in the impurity dynamics, as we will elaborate in this paper.

 In this work, we will also consider another factor which influences the impurity dynamics, i.e., the external confinement of bosons. Such a confining potential gives rise to an inhomogeneous boson density distribution, which leads to an additional potential for the impurity via the impurity-boson interactions. In fact, it is such a harmonic confinement that breaks the translational invariance of the whole system, while its effect on the impurity dynamics has not been considered in previous studies~\cite{expt, Gangardt1, Gangardt2, Gangardt3}.  Here we explicitly relate the two effective potentials, i.e., the periodic and the harmonic ones, to the coupling strength between the impurity and bosons. We show that the period and amplitude of the impurity dynamics sensitively depend on the sign and strength of the impurity-boson coupling.
Our results offer insights into the impurity dynamics under a general class of background systems as long as they are fermionized.

 The rest of the paper is organized as follows. In section II, we set up the basic model for the impurity moving in the background of hardcore bosons. Based on the model, we study the impurity dynamics under an external force in section III. Section IV is contributed to the discussion and summary of our results.

\section{Model}

We start from the Hamiltonian of the system following the setup in Ref.~\cite{expt} ($\hbar=1$ throughout the paper):
\begin{equation}
H=H_{b}(x_1,\cdots,x_N)-\frac{1}{2m}\frac{\partial^2}{\partial x^2}+g_{ib}\sum_{i=1}^N\delta(x-x_i)-F x.
\end{equation}
Here $x$ and $x_{i=1,2...N}$ are respectively the coordinates of impurity and N bosons; $g_{ib}$ is the impurity-boson coupling strength; $F$ is the external force acting on the impurity starting from time $t=0^+$; $H_b$ is the Hamiltonian for the hard-core bosons:
\begin{equation}
H_{b}=\sum_i\bigg(-\frac{1}{2m}\frac{\partial^2}{\partial x_i^2}+\frac{1}{2}m\omega_{ho} x_i^2\bigg)+g_{bb}\sum_{i<j}\delta(x_i-x_j).\label{Ham}
\end{equation}
In the limit $g_{bb}\rightarrow\infty$, the ground state of bosons can be written as $\Psi_b(x_1,...,x_N)=|\phi_F(x_1,...,x_N)|$, where $\phi_F$ is the Slater determinant describing N identical fermions occupying the lowest N levels of an 1D harmonic oscillator. As the spatial order of impenetrable particles is fixed, one can write down the probability density of finding the i-th ordered particle at $x$, named as $\rho_i(x)$, to be
\begin{equation}
\rho_i(x)=\int d\vec{x} |\phi_F|^2 \theta(x_1<...<x_i<...<x_N) \delta(x-x_i),
\end{equation}
It has been shown that each $\rho_i(x)$ is well separated and follows a Gaussian distribution centered at $\bar{x}_i=\int dx x \rho_i(x)$ and with a width $\sigma_i$~\cite{Yang}:
\begin{equation}
\rho_i(x)\rightarrow \frac{1}{\sqrt{\pi}\sigma_i}e^{-(x-\bar x_i)^2/\sigma_i^2}. \label{rho_i}
\end{equation}
By mapping the order index to site index, one can consider each ordered particle as localized in the corresponding lattice site with a finite distribution width. Following this idea one can construct various spin chain models as studied in the literature~\cite{Santos, Zinner, Pu, Levinsen,Yang, Cui, Blume, Zinner2, Levinsen2, Cui2, Chen, Pu2, Santos2}.

Assuming that the impurity-boson interaction is weak enough compared to the Fermi energy of the hard-core bosons, i.e., $g_{ib}/d\ll E_F=N\omega_{ho}$ ($d$ is inter-particle distance of bosons), the ground-state profile of bosons will not be significantly changed by the impurity. In this case, we can
write down an effective potential for the impurity due to impurity-boson interaction:
\begin{eqnarray}
V(x)=g_{ib}\sum_i \rho_i(x). \label{V_imp}
\end{eqnarray}
Therefore the ``lattice" structure of $\rho_i(x)$ is naturally transferred to a ``lattice" potential on the impurity. The resulting Hamiltonian for the impurity is
\begin{equation}
H_{imp}=-\frac{1}{2m}\frac{\partial^2}{\partial x^2}+V(x)-F x. \label{H_imp}
\end{equation}

In the ideal situation when $\rho_i(x)$ (Eq.\ref{rho_i}) is equally distributed with the same width, i.e., $\bar x_{i+1}-\bar x_i\equiv d, \sigma_i\equiv \sigma$, $V$ reduces to the ideal lattice potential $V_{L}$ for large $N$:
\begin{equation}
V_{L}(x)=g_{ib}\sum_i \frac{1}{\sqrt{\pi}\sigma}e^{-(x-\bar{x}_i)^2/\sigma^2}, \ \ \ \bar{x}_i=(i-\frac{N-1}{2})d\label{V_L}
\end{equation}
which is translationally invariant and can support BO dynamics of the impurity under an external force. Remarkably, here the ``lattice" is induced by the finite impurity-boson coupling $g_{ib}$, and therefore the property of BO is highly tunable by $g_{ib}$. This is the unique aspect of such an interaction-induced BO.


Meanwhile, it should be noted that one essential deviation between $V_L$ and actual $V$ is because of the inhomogeneity of boson density in a trap. In particular, the height of $\rho_{i}$ changes with index $i$, which decays gradually from the trap center to the edge. This generates an additional potential $V'(x)$ on top of $V_L$, as shown by dashed line in Fig.~1. $V'$ can be estimated through the local density approximation (LDA), and for small $x$ it is
\begin{equation}
V'(x)\sim -sgn(g_{ib})\frac{1}{2}m\omega'^2 x^2, \ \ \ \ \omega'=\sqrt{\frac{|g_{ib}|\omega_{ho}}{\pi R}}, \label{V'}
\end{equation}
here $R=(2N)^{1/2} a_{ho}\ (a_{ho}=1/\sqrt{m\omega_{ho}})$ is the Thomas radius of hardcore bosons under LDA. Eq.~\ref{V'} shows that $V'$ is simply a harmonic potential with the frequency scaling with $|g_{ib}|^{1/2}$. A subtle case is that when $g_{ib}$ is repulsive and $V'$ is concave, one would have to impose another harmonic potential to host the impurity initially at the trap center. We will discuss this case later.

Now we can approximate $V$(Eq.~\ref{V_imp}) as the sum of a lattice potential $V_L$(Eq.\ref{V_L}) and a harmonic one $V'$(Eq.\ref{V'}). Such an approximation is expected to work well for the impurity dynamics near the center of the trap, but not near the edges where the assumption of uniform Gaussian distribution in $V_L$ breaks down. The individual effect of $V_L$ and $V'$ to the impurity dynamics is analyzed as follows. $V_L$ induces BO with the period $2\pi/(Fd)$ and an amplitude
proportional to the band width, which is a decreasing function of
$|g_{ib}|$; $V'$ also induces a periodic dynamics due to the linear
interference between different harmonic levels, and the resulting
period and amplitude of the oscillation all depend on $\omega'$ or
$g_{ib}$ (see Eq.\ref{V'}). So under the combined effects of $V_L$
and $V'$, the impurity is expected to undergo oscillatory dynamics
with properties crucially relying on the coupling $g_{ib}$.

\section{Results}

In our numerical simulation of the impurity dynamics under $H_{imp}$ (Eq.\ref{H_imp}), we have chosen the initial state as the ground state of $H_{imp}$ without external force ($F=0$). We then turn on a finite $F$ at time $t=0^+$ and solve the time-dependent Schr\"odinger equation $i\frac{\partial\psi}{\partial t}= H_{imp}\psi$, with $\psi$ the impurity
wave function. In the simulation, we have discretized the coordinate space in a sufficiently large region (with the size much larger than the Thomas radius $R$) in order to numerically solve the ground state and the dynamics. Here we define the dimensionless parameters
$\tilde{g}_{ib}\equiv g_{ib}/(\omega_{ho} a_{ho})$, and $\tilde{F}\equiv Fa_{ho}/\omega_{ho}$.

\begin{figure}[h]
\includegraphics[height=3.5cm,width=9cm]{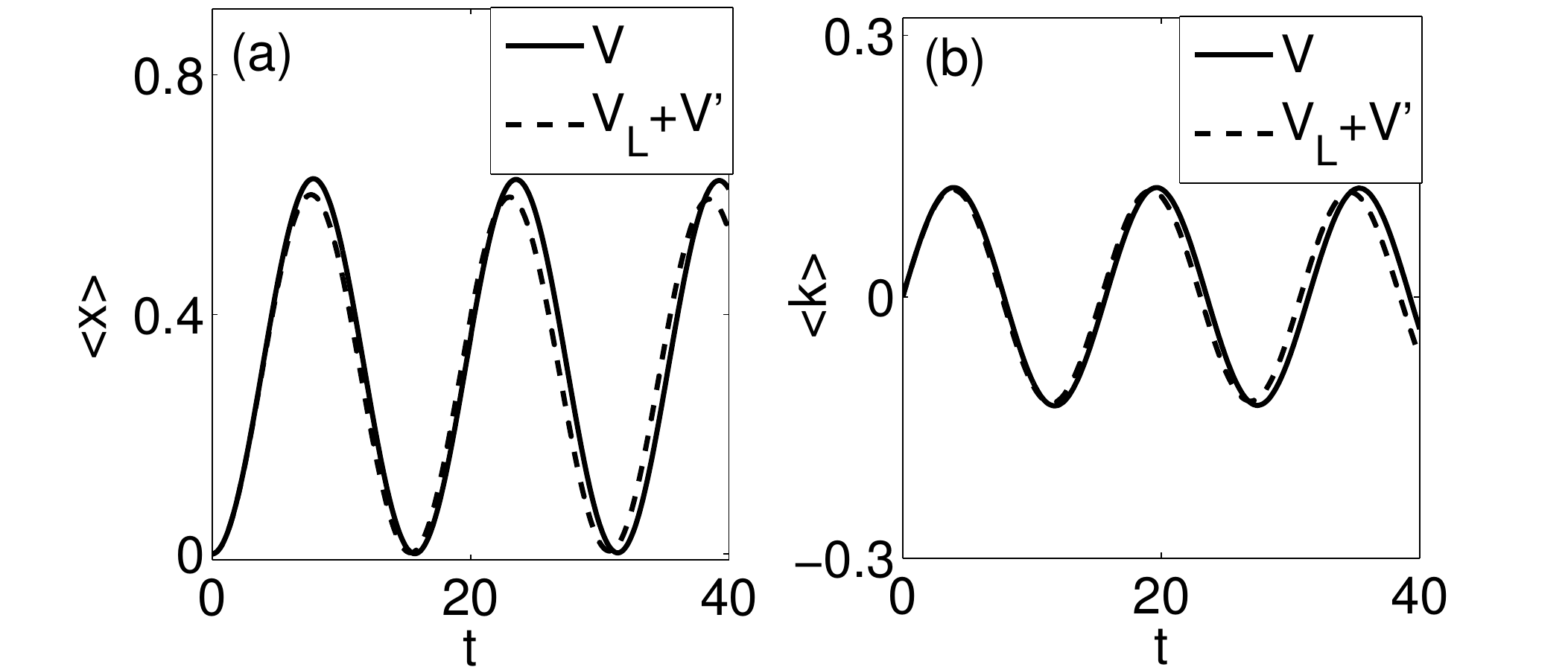}
\caption{Mean displacement $\langle x\rangle$ (a) and mean momentum $\langle k\rangle$(b) of the impurity as time evolves. Here $\tilde{g}_{ib}=-2$, $\tilde{F}=0.05$. The exact results (solid lines, using potential $V$ (Eq.\ref{V_imp})) are compared with those by replacing $V$ with $V_L+V'$ (dashed).
Here the units of length, momentum and time are respectively $a_{ho},\ 1/a_{ho}$ and $1/\omega_{ho}$.} \label{fig2}
\end{figure}

In Fig.~\ref{fig2}, we plot the time evolution of the mean displacement $\langle
x\rangle\equiv \langle \psi |x|\psi\rangle$ and the mean momentum $\langle k\rangle\equiv \langle \psi |-i\frac{\partial}{\partial x} |\psi\rangle$ for the impurity
moving in the background of $N=10$ bosons, taking
$\tilde{g}_{ib}=-2$ and $\tilde{F}=0.05$ for instance. In this case,
the resulting $\omega'=0.37\omega_{ho}$ according to Eq.~\ref{V'}.
As expected, both $\langle x\rangle$ and $\langle k\rangle$
oscillate periodically in time $t$, and the exact results (by
simulating Eq.~\ref{H_imp}) can be fitted quantitatively well by
replacing the potential $(V)$ with the sum of lattice and harmonic
potentials $(V_L+V')$.

\begin{figure}[h]
\includegraphics[height=6cm,width=8.5cm]{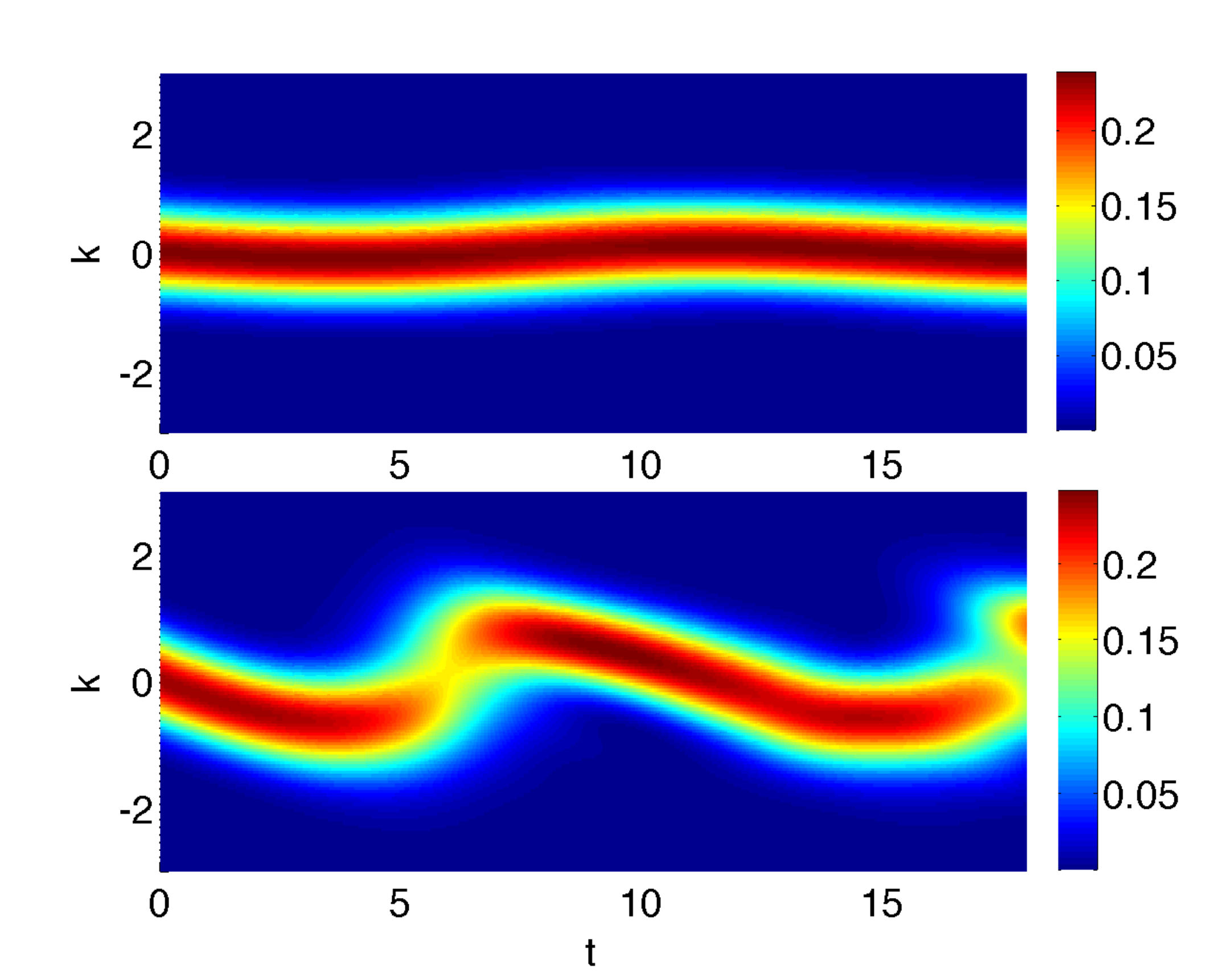}
\caption{(Color online). Momentum distribution $n(k)$ of the impurity in the parameter plane of momentum $k$ and time $t$. Here $\tilde{g}_{ib}=-2$; the force $\tilde{F}=0.05$ in (a) and $0.3$ in (b). The units are the same as used in Fig.~\ref{fig2}.} \label{fig_new}
\end{figure}

In Fig.~\ref{fig_new}, we present the impurity momentum distribution, $n(k)\equiv \int dx e^{ik(x-x')}\psi^*(x)\psi(x')$, in the parameter plane of momentum $k$ and time $t$. In general, the behavior of $n(k)$ will depend on the strengths of impurity-boson coupling $g_{ib}$ and external force $F$. To see the effect of the external force, we choose a small $\tilde{F}=0.05$ in Fig.~\ref{fig_new}(a) and a larger one $\tilde{F}=0.3$ in Fig.~\ref{fig_new}(b). We see that the sharp Bragg reflection, as has been discussed in Ref.~\cite{expt}, is more visible for larger $\tilde{F}$ (Fig.~\ref{fig_new}(b)). As the harmonic potential $V'$ can only give rise to a periodic oscillation of $n(k)$, such a reflection can only be attributed to the effect of the underlying periodic potential $V_L$, or the hidden "lattice" structure of hardcore bosons.

Nevertheless, here we would like to point out that a sharp reflection of momentum in $n(k)$ can be a sufficient condition, but not a necessary condition, for the BO dynamics under the periodic potential. As observed earlier in the optical lattice experiment~\cite{BO_OL1}, such a reflection in $n(k)$ is only visible for shallow lattices, but not for deep ones. This is because for deep lattices, each crystal momentum state is the superposition of many plane-wave states, and the time evolution of these (plane-wave) momenta will all contribute to $n(k)$. This will result in a perfect periodic oscillation of $n(k)$ as shown in Ref.~\cite{BO_OL1}.

\begin{figure}[h]
\includegraphics[width=9cm]{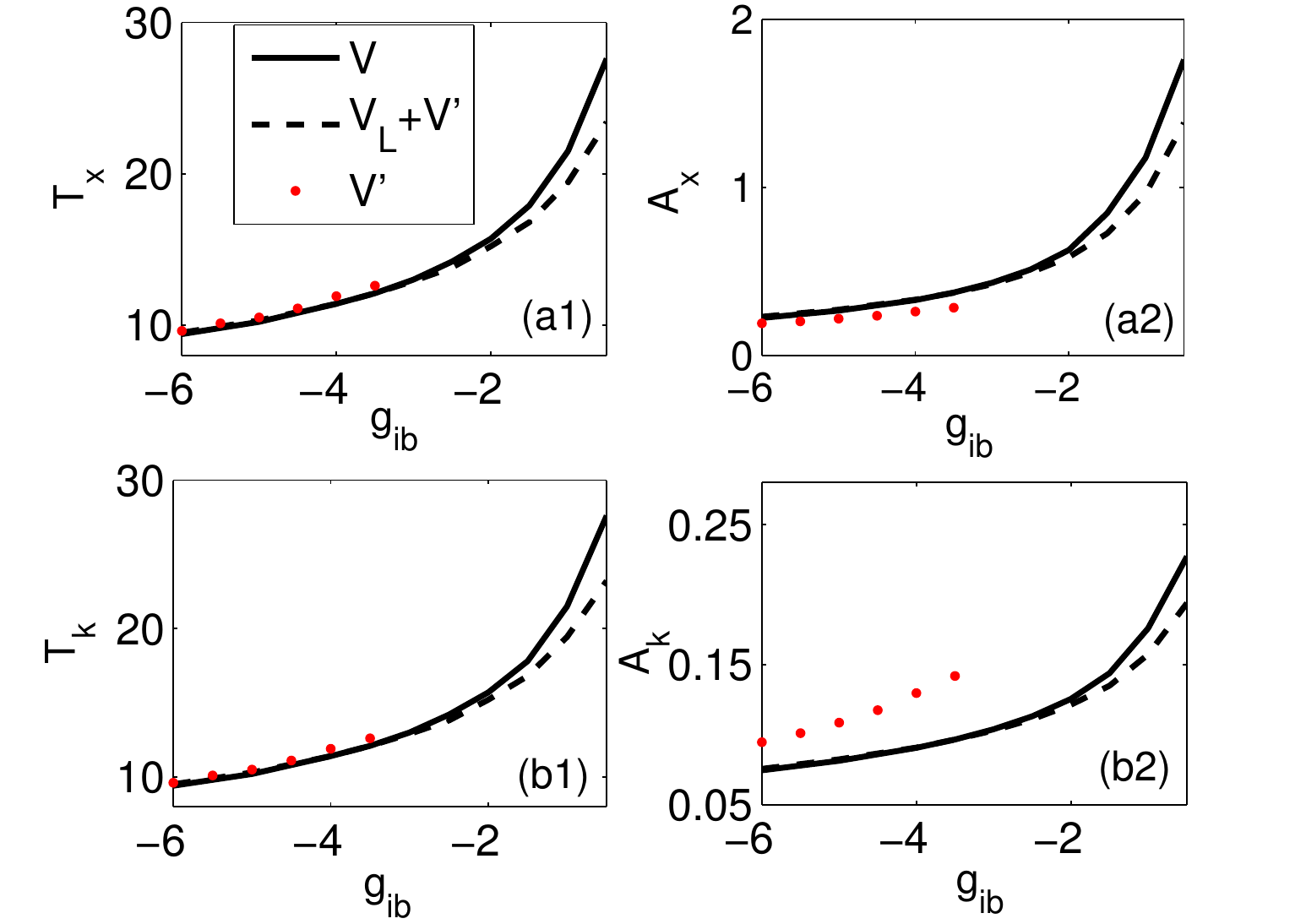}
\caption{(Color online). (a1,a2): the period and amplitude of $\langle x\rangle $ as functions of attractive $g_{ib}$. (b1,b2): the period and amplitude of $\langle k\rangle $ as functions of attractive $g_{ib}$. Results from different models are compared.  Here $\tilde{F}=0.05$, and $g_{ib}$ is in the unit of $(\omega_{ho} a_{ho})$. Other units are the same as in the caption of Fig.~\ref{fig2}.
} \label{fig3}
\end{figure}

To see clearly the role of $g_{ib}$ in the dynamics, we
extract the period $T_x$ ($T_k$) and amplitude $A_x$ ($A_k$) for
$\langle x\rangle $ ($\langle k\rangle$) and plot them as functions
of $g_{ib}$ in Fig.~\ref{fig3}($g_{ib}<0$) and Fig.~\ref{fig3}($g_{ib}>0$). Again we see
that the results from periodic and harmonic potentials $(V_L+V')$ fit well with
exact results from the total potential $(V)$.

For the attractive $g_{ib}$ case in Fig.~\ref{fig3}, we see that all the periods ($T_x,T_k$) and amplitudes ($A_x,A_k$) decrease as $|g_{ib}|$. This can be attributed to the two effects generated by increasing $|g_{ib}|$. First, it will deepen the lattice depth of $V_L$ and produce a narrower band width, which reduces the amplitudes $A_x$ and $A_k$. Second, it will produce a tighter confinement $V'$(see Eq.~\ref{V'}), which further reduces $A_x$ and $A_k$ as well as $T_x$ and $T_k$. In fact, as $|g_{ib}|$ becomes larger, the effect of $V'$ will dominate and the dynamics essentially follow the harmonic prediction (red-dotted lines in Fig.~3).

\begin{figure}[h]
\includegraphics[width=9cm]{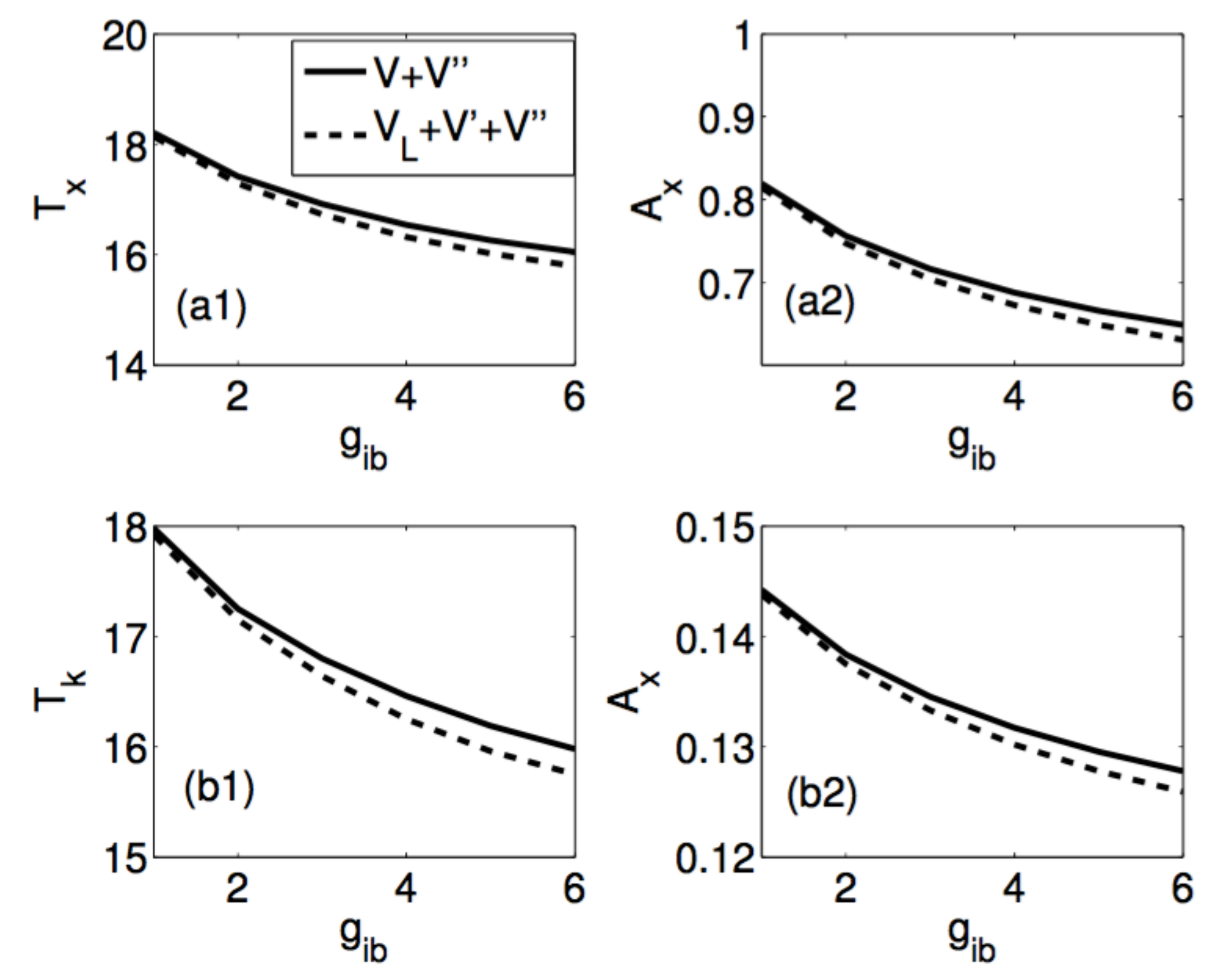}
\caption{Same to Fig.3 but for repulsive $g_{ib}$. In order to compensate the concave potential $V'(x)$, we activate an additional (convex) harmonic potential $V''(x)$ for the impurity with frequency $\omega''=\omega'+0.2\omega_{ho}$. } \label{fig4}
\end{figure}

For repulsive $g_{ib}$, as discussed before, one has to
impose another harmonic potential, $V''(x)=\frac12 m\omega''^2 x^2$, to
compensate for the effect of concave potential $V'$ (Eq.~\ref{V'})
and host the impurity initially at trap center ($\langle x\rangle_{t=0}=0$). In order to highlight the role of $V_L$ in the
dynamics, we have chosen $\omega''$ just a bit larger than $\omega'$, i.e., $\omega''=\omega'+0.2\omega_{ho}$, and the results are shown in Fig.~\ref{fig4}. We see that as $g_{ib}$ increases,
$A_x$, $A_k$ and $T_x$, $T_k$ all decrease. Similar to $g_{ib}<0$ case, this can be attributed to the narrower band width produced by larger $g_{ib}$ (and thus deeper $V_L$), as well as the combined effects of $V_L$ and residue harmonic potential $V'+V''$.

Actually, by expanding $H_{imp}$(Eq.\ref{H_imp}) in terms of the
lowest-band Wannier functions that are supported by $V_L$, we can write
down an effective lattice model for the impurity:
\begin{equation}
H_{imp}^{eff}=-\sum_{(i,j)}t_{ij}(c_i^\dagger c_j+h.c.)+ \sum_i
(V^h_i-F_i) c_i^\dagger c_i, \label{H_eff}
\end{equation}
here $V^h_i$ and $F_i$ are respectively the on-site potential
generated by the total harmonic confinement (sum of $V'$ and $V''$)
and the force $Fx$; the hopping is
\begin{eqnarray}
t_{ij}&=&-\int w_0^*(x)\left(-\frac{1}{2m}\frac{\partial^2}{\partial
x^2}+V_L(x)\right)w_0(x-(j-i)d)dx. \nonumber
\end{eqnarray}
Note that the lattice model (\ref{H_eff}) is valid under the
adiabaticity condition~\cite{BO_OL1}, which requires $Fd\ll E_{gap}$
($E_{gap}$ is the band gap) to ensure the dynamics within the lowest
band. Apparently this condition is satisfied for large $|g_{ib}|$ (and thus large band gap). We have checked that it is not satisfied by for the range of $g_{ib}$ considered in
Fig.~\ref{fig3} and Fig.~\ref{fig4}, where the higher band effects should play essential roles in determining the dynamics.

\section{Discussion and summary}

Our results offer a number of insights into the oscillatory impurity
dynamics in Ref.~\cite{expt}. First, such dynamics is purely induced
by the impurity-boson interaction $g_{ib}$. Increasing $|g_{ib}|$
will generally lead to faster oscillatory dynamics with smaller
amplitudes, which is qualitatively consistent with what was observed in
Ref.~\cite{expt}. Second, both the induced ``lattice" potential
(due to the ``lattice" structure of bosons) and the harmonic
confinement (due to inhomogeneous boson density) play important
roles in the resulting dynamics. Their individual effects can be
examined by tuning $g_{ib}$ (repulsive or attractive) or applying
additional confinements on the impurity. Third, the impurity
dynamics does not rely on the statistics of the background system, but rather on the fact that the system is fermionized. Physically, this is because any fermionized system has the same density profile, regardless of whether it is composed of bosons, fermions or boson-fermion mixtures. Fermionized backgrounds thus affect the impurity similarly via density-density interactions.

It is worthwhile to point out that the periodic dynamics in this work is
related to the assumption of unaffected boson profile.
Once the coupling $g_{ib}$ is strong enough to invalidate this
assumption, the boson excitation should be taken into account, which
is expected to bring more modes into the dynamics and cause
damping, as observed in the Innsbruck experiment~\cite{expt}. The study
of dynamics in this regime is beyond the scope of the present work.

In summary, we have demonstrated the interaction-induced oscillatory
dynamics of an impurity moving in the background of an 1D trapped
hard-core bosons. Because of the hidden ``lattice" structure of
bosons, the impurity dynamics essentially mimics the BO in conventional
lattices, despite the lack of lattice translational
invariance. Moreover, we also point out that the inhomogeneous
density of trapped bosons provides another harmonic potential that
can strongly affect the dynamics. These results provide a new
perspective on the recent observation in the Innsbruck
experiment~\cite{expt}.

{\bf Acknowledgment.}
This work is supported by the National Natural Science Foundation of China (No.11374177, 11374283, 11626436, 11421092, 11534014, 11522545), and the National Key Research and Development Program of China (No.2016YFA0300603,  2016YFA0301700). W. Y. acknowledges support from the ``Strategic Priority Research Program(B)'' of the Chinese Academy of Sciences, Grant No. XDB01030200.

{\bf Note Added}: During preparing this paper, we became aware of the preprint by Yang and Pu~\cite{Pu3}, who studied the impurity dynamics in a different interaction regime ($g_{ib}\rightarrow+\infty$).

\end{document}